\documentclass[amsmath,amssymb,prl,hyperlink,twocolumn]{revtex4}
\usepackage[paperwidth=210mm,paperheight=297mm,centering,hmargin=2.0cm,vmargin=2.6cm]{geometry}

\usepackage{graphicx}
\usepackage{soul}
\usepackage[colorlinks=true,citecolor=blue,linkcolor=magenta]{hyperref}
\usepackage[usenames]{color}
\usepackage{amsfonts}

\begin{document}
\title{Temperature-dependent Mollow triplet spectra from a single quantum dot: \\ Rabi frequency renormalisation and sideband linewidth insensitivity}
\author{Yu-Jia Wei}
\affiliation{Hefei National Laboratory for Physical Sciences at the Microscale and Department of Modern Physics, \& CAS Center for Excellence and Synergetic Innovation Center in Quantum Information and Quantum Physics, University of Science and Technology of China, Hefei, Anhui 230026, China}
\author{Yu He}
\affiliation{Hefei National Laboratory for Physical Sciences at the Microscale and Department of Modern Physics, \& CAS Center for Excellence and Synergetic Innovation Center in Quantum Information and Quantum Physics, University of Science and Technology of China, Hefei, Anhui 230026, China}
\author{Yu-Ming He}
\affiliation{Hefei National Laboratory for Physical Sciences at the Microscale and Department of Modern Physics, \& CAS Center for Excellence and Synergetic Innovation Center in Quantum Information and Quantum Physics, University of Science and Technology of China, Hefei, Anhui 230026, China}
\author{Chao-Yang Lu}
\affiliation{Hefei National Laboratory for Physical Sciences at the Microscale and Department of Modern Physics, \& CAS Center for Excellence and Synergetic Innovation Center in Quantum Information and Quantum Physics, University of Science and Technology of China, Hefei, Anhui 230026, China}
\author{Jian-Wei Pan}
\affiliation{Hefei National Laboratory for Physical Sciences at the Microscale and Department of Modern Physics, \& CAS Center for Excellence and Synergetic Innovation Center in Quantum Information and Quantum Physics, University of Science and Technology of China, Hefei, Anhui 230026, China}
\author{Christian Schneider}
\affiliation{Technische Physik, Physikalisches Institut and Wilhelm Conrad Rš\"ontgen-Center for Complex Material Systems, Universit\"aŠt W\"uŸrzburg, Am Hubland, D-97074 W\"uŸrzburg, Germany}
\author{Martin Kamp}
\affiliation{Technische Physik, Physikalisches Institut and Wilhelm Conrad Rš\"ontgen-Center for Complex Material Systems, Universit\"aŠt W\"uŸrzburg, Am Hubland, D-97074 W\"uŸrzburg, Germany}
\author{Sven H\"ofling}
\affiliation{SUPA, School of Physics and Astronomy, University of St. Andrews, St. Andrews KY16 9SS, United Kingdom}
\affiliation{Technische Physik, Physikalisches Institut and Wilhelm Conrad Rš\"ontgen-Center for Complex Material Systems, Universit\"aŠt W\"uŸrzburg, Am Hubland, D-97074 W\"uŸrzburg, Germany}
\affiliation{Hefei National Laboratory for Physical Sciences at the Microscale and Department of Modern Physics, University of Science and Technology of China, Hefei, Anhui 230026, China}
\author{Dara P. S. McCutcheon}\email{daramc@fotonik.dtu.dk}
\address{DTU Fotonik, Department of Photonics Engineering, {\O}˜rsteds Plads, 2800 Kgs Lyngby, Denmark}
\affiliation{{Departamento de F\'isica, FCEyN, UBA and IFIBA, Conicet, Pabell\'on 1, Ciudad Universitaria, 1428 Buenos Aires, Argentina}}
\author{Ahsan Nazir}\email{ahsan.nazir@manchester.ac.uk}
\affiliation{Photon Science Institute \& School of Physics and Astronomy, University of Manchester, Oxford Road, Manchester M13 9PL, United Kingdom}
\affiliation{Controlled Quantum Dynamics Theory, Imperial College London, London SW7 2AZ, United Kingdom}

\date{\today}

\begin{abstract}
We investigate temperature-dependent 
resonance fluorescence spectra obtained from a single self-assembled quantum dot. 
A decrease of the Mollow triplet sideband splitting is observed with increasing temperature, an effect we attribute to a
phonon-induced renormalisation of the driven dot Rabi frequency. We also present first evidence for a non-perturbative regime 
of phonon coupling, in which the expected linear increase in sideband linewidth as a function of temperature is cancelled by the 
corresponding reduction in Rabi frequency. These results indicate that dephasing in semiconductor quantum dots may be less 
sensitive to changes in temperature than expected from a standard weak-coupling analysis of phonon effects.

\end{abstract}

\maketitle

Self-assembled semiconductor quantum dots (QDs) provide a promising platform for quantum information processing
using single spins~\cite{spins,spins2} and photons~\cite{photons1,photons2,photons3,photons4}. 
Such applications require QD quantum coherence to be preserved on timescales sufficient
for performing high fidelity quantum operations, and for emitted single photons  
to possess a large degree of indistinguishability~\cite{flagg10,patel10,nazir09b}.  
Indistinguishable photons can be produced by s-shell resonant optical excitation of a single QD~\cite{vamivakas09,flagg09,muller07},
wherein an electron-hole pair is created directly without any relaxation from higher states, which would otherwise cause inhomogeneous broadening in the QD emission spectrum. The coherence time ($T_2$) of such photons is able to approach the Fourier transform limit, $T_2=2T_1$ (with $T_1$ the QD radiative lifetime) at low temperatures ($\sim4$~K) and weak driving strengths~\cite{matthiesen12}.

As the intensity of the pump laser increases, however, an additional power dependent dephasing contribution arises, 
even at low temperatures~\cite{ramsay10,ramsay10_2,ulrich11_short,monniello13,forstner03,machnikowski04,vagov2007_short,mogilevtsev08,nazir08,mccutcheon10_2,kim13}.
This is often termed excitation induced dephasing (EID), which commonly 
originates from deformation potential coupling of QD excitons to longitudinal acoustic (LA)
phonons. As the driving strength increases, so does the energy splitting of the excitonic dressed states, 
and excitations are then able to scatter with the increased density of phonons 
around this energy scale in the bulk semiconductor lattice. Driving dependence
is thus a pronounced characteristic of EID, as was observed in Refs.~\cite{ramsay10,ramsay10_2}, where QD excitonic Rabi oscillations
were measured via photocurrents, and in Refs.~\cite{ulrich11_short,monniello13} through driven QD optical emission. 

Besides EID, another direct manifestation 
of the phonon influence on a driven dot can be found in the dependence of its properties 
on {\emph{temperature}}. In fact, an idealised two-level-system (e.g.~an isolated atom) should not show any change in emission behaviour
with temperature over the usual experimental range,
in contrast to the response to changes in driving strength, which is nontrivial even in the
absence of phonons~\cite{mollow69}. Though the damping of excitonic Rabi rotations (in the QD population) has 
been measured at various temperatures under pulsed excitation~\cite{ramsay10,ramsay10_2,monniello13}, 
the direct observation of temperature dependence in the field correlation properties and resonance fluorescence (RF) \emph{emission spectra} of a
single QD has not been reported. 
This is arguably the most relevant scenario for photonic applications, since it is crucial to understand the competing roles
played by the electromagnetic and solid-state environments in determining the QD optical emission characteristics and coherence~\cite{mccutcheon13,roy11,roy12,moelbjerg12}.

Here, we present a combined experimental and theoretical investigation 
of temperature dependence in the RF 
spectra obtained from a single self-assembled QD,
generated by continuous-wave (cw) s-shell excitation. 
We show that the coupling of the QD 
neutral exciton to LA phonons leads to a temperature-dependent phonon renormalisation (suppression) of the driven dot Rabi frequency,
which we can observe directly through a decrease in Mollow triplet sideband splitting with increasing temperature. This trend
is in agreement with an exciton-phonon coupling model 
based on a polaron master equation technique~\cite{mccutcheon10_2,roy11,roy12,mccutcheon13}.
Our theoretical analysis also predicts a non-perturbative regime of phonon-coupling, wherein a linear increase of the sideband linewidths with temperature (as would be found 
from a standard weak-coupling analysis~\cite{ramsay10}) is appreciably cancelled 
by the corresponding decrease in the phonon-renormalised 
Rabi frequency~\cite{mccutcheon10_2}. This prediction is supported by our experimental data, demonstrating
an experimental regime in which EID is less sensitive to changes in temperature than would be expected 
from a perturbative, weak exciton-phonon coupling model~\cite{ramsay10,ramsay10_2,nazir08,machnikowski04}.

\begin{figure}
\centering
\includegraphics[width=0.48\textwidth]{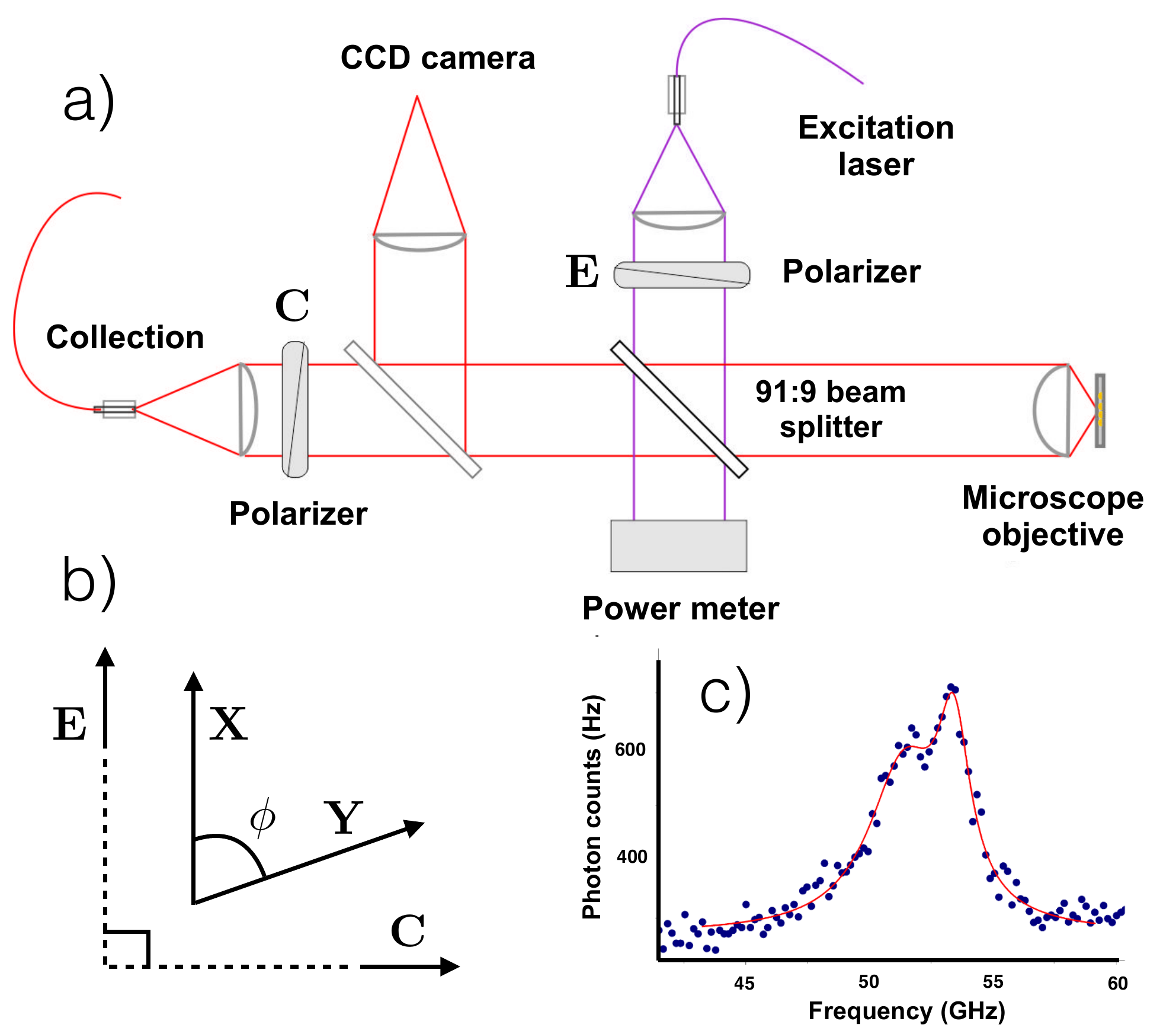}
\caption{The cross-polarization experimental setup is shown in a), where vectors associated with 
the polarizers in the excitation {\bf{E}}, and collection {\bf{C}} arms are approximately orthogonal, as 
shown in b). Also shown schematically in b) are the dipole moment orientation vectors of the two fine-structure split states 
{\bf{X}} and {\bf{Y}}. These states are split by approximately $2~\mathrm{GHz}$ as confirmed 
by above band photo-luminescence measurements shown in c).
}
\label{setup}
\end{figure}

Our experiments are performed on a single self-assembled InAs/GaAs QD embedded in the centre of a micro-cavity 
with a low Q factor ($\sim 200$) in order to enhance fluorescence
collection efficiency. The sample is kept in a cryogen-free bath cryostat 
with the temperature stabilised within $\pm$5~mK around a fixed value, as measured by a thermometer on the sample 
holder~\cite{note}. 
A cw pump is provided by a semiconductor diode laser, with frequency stabilised 
to within $\pm2\pi\times2$~MHz, which is much narrower than the linewidth of the QD fluorescence. 
Fig.~\ref{setup} shows the cross-polarization setup used to achieve resonant excitation~\cite{matthiesen12}. 
The neutral single exciton of the QD has two fine-structure states, whose dipole moment orientations 
we label {\bf{X}} and {\bf{Y}}. As seen in Fig.~\ref{setup}c), these states are split by approximately $2~\mathrm{GHz}$. 
We resonantly excite the {\bf{Y}} transition with a polarization approximately orthogonal to the collection polarization. 
Since {\bf{X}} and {\bf{Y}} are not strictly orthogonal, and the {\bf{X}} transition 
is detuned and orthogonal to the collection polarisation {\bf C}, we can selectively measure RF emission 
from the {\bf{Y}} transition while extinguishing the pump laser with a ratio exceeding $10^6$. Non-orthogonality 
of the fine structure states may be due to imperfect selection rules within the solid-state, and has been reported elsewhere~\cite{santori01}. 
Finally, a Fabry-Perot (FP) interferometer is used to resolve the RF spectra with a spectral resolution of 
$\sim2\pi\times67$~MHz (which we account for in our fittings below).

We begin by analysing QD RF spectra at \emph{fixed temperature} ($4$~K), for a series of cw 
laser driving strengths~\cite{photons3}. 
This serves as a form of calibration of our system, and through a comparison with theory
provides us with necessary parameters for our subsequent examination 
of temperature changes in the spectra. High resolution RF spectra from the neutral exciton state 
were obtained for various laser powers~\cite{photons3}. 
For increasing driving strength, the expected triple peak structure in the RF spectra, known as the Mollow triplet~\cite{mollow69}, was observed.
This can be interpreted in the dressed state picture~\cite{cohenbook}, 
in which the QD exciton is dressed through its interaction with the driving laser. 
This causes the originally degenerate, uncoupled QD-photon states to split into two states  
separated by $\Omega_r$, where $\Omega_r$ is the effective phonon-renormalised strength of the QD-laser interaction.  
Photons can then be emitted corresponding to four different transitions, two at the QD transition frequency, and two more offset by 
$\pm\Omega_r$. In the absence of phonons, and within the dipole
approximation, $\Omega_r\to\Omega=\mu E/\hbar$ is just the bare Rabi frequency, where
$\mu$ is the effective dot dipole moment and $E$ is the amplitude of optical field at the QD.

The presence of phonons, however, introduces both additional broadening and a renormalisation of the effective Rabi strength, $\Omega\to\Omega_r$.
To account for these processes, we exploit a polaron master equation 
technique~\cite{mccutcheon10_2,roy11,roy12,mccutcheon13}, which allows us to derive
an explicit expression for the emission spectrum of the QD in terms of experimental 
parameters and microscopic constants characterising the QD-phonon coupling.
We model the excitonic degree of freedom of the QD as a two-level-system, and couple it to two harmonic oscillator baths,
which represent both the electromagnetic field into which light is emitted, and the phonon environment present in the substrate.
In Ref.~\cite{mccutcheon13} it was 
shown that provided $\Omega<k_B T<\omega_c$, with $T$ the temperature and $\omega_c$ the cut-off frequency of the
phonon environment (related to the size of the QD~\cite{ramsay10_2}), the influence of phonons on the incoherent emission spectrum can be 
captured entirely by a renormalisation of the
bare Rabi frequency and the introduction of a pure-dephasing rate. 
Above saturation, where $\Omega_r\gg \Gamma_1$, 
with $\Gamma_1=1/T_1$ and $T_1=390\pm10$~ps determined by time resolved correlation measurements, 
the emission spectrum is given by
$S(\omega)\propto\mathrm{Re}[\int_{0}^{\infty}\mathrm{d}\tau\mathrm{e}^{i(\omega-\omega_l)\tau}g_1(\tau)]$, 
in terms of the first order field correlation function 
\begin{equation}\label{g1approx}
g_1(\tau)\approx\frac{1}{4}\left(e^{-\Gamma_2\tau}+e^{-(1/2)(\Gamma_1+\Gamma_2)\tau}\cos(\Omega_r \tau)\right),
\end{equation}
with $\omega_l$ the frequency of the excitation laser. The renormalised Rabi frequency is (we set $\hbar=1$)
\begin{align}
\label{Omegar}
\Omega_r(T)=\Omega \exp{\left[-{\textstyle{\frac{1}{2}}}\int_0^{\infty}\mathrm{d}\omega\frac{J(\omega)}{\omega^2}\coth\Big(\frac{\omega}{2k_B T}\Big)\right]},
\end{align}
where $J(\omega)$ is the spectral density of the QD-phonon interaction, which for coupling to LA phonons has been shown to be adequately described 
by the functional form $J(\omega)=\alpha \,\omega^3 \exp[-(\omega/\omega_c)^2]$, with $\alpha$ capturing the strength of the interaction~\cite{ramsay10,ramsay10_2}.
The phonon-induced pure-dephasing rate enters through $\Gamma_2=(1/2)\Gamma_1+\gamma_{\mathrm{PD}}+\gamma_0$, and
is given by 
\begin{equation}\label{gammapd}
\gamma_{\rm PD}=\left(\frac{\Omega_r}{2}\right)^2\int_{-\infty}^{\infty}\mathrm{d}s\cos{(\Omega_rs)}(e^{\phi(s)}-e^{-\phi(s)}),
\end{equation}
with $\phi(s)=\int_0^{\infty}\mathrm{d}\omega J(\omega)\omega^{-2}[\cos{(\omega s)}\coth{(\omega/2k_BT)}-i\sin{(\omega s)}]$. We also include
a term $\gamma_0$ to account for any additional {\it driving-independent} pure-dephasing processes.

\begin{figure}
\centering
\includegraphics[width=0.48\textwidth]{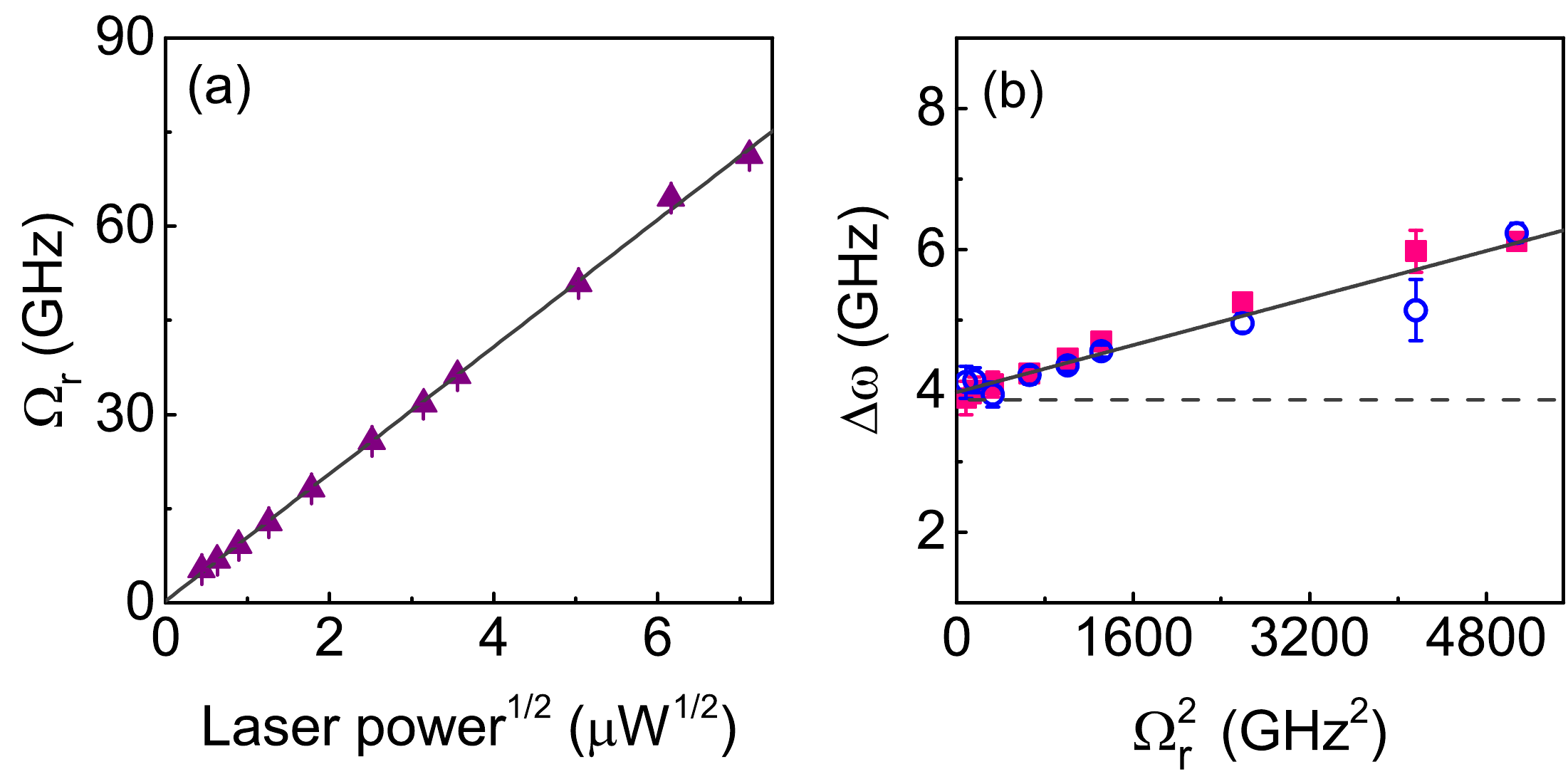}
\caption{(a) Extracted phonon renormalised Rabi frequency $\Omega_r$ as a function of the square root of the cw laser power
$\sqrt{P}$, together with a linear fit (grey solid line).
(b) Extracted linewidths of the red (red filled squares)
and blue (blue open circles) sidebands
as a function of $\Omega^2_r$. A straight line fit to the data points is shown in 
grey (solid line). 
The dashed line shows the expected linewidth in the
absence of any pure dephasing (driving-dependent or otherwise), i.e.~the Fourier transform limit.}
\label{driving}
\end{figure}

In the regime described above, the spectrum is given simply by the sum of three Lorenztians corresponding
to the three peaks in the Mollow triplet. 
We are interested here in the sidebands, 
which have positions at $\pm\Omega_r$ and a
full-width-half-maximum (FWHM) of 
\begin{equation}\label{fwhm}
\Delta\omega=\Gamma_1+\Gamma_2=(3/2)\Gamma_1+\gamma_{\mathrm{PD}}+\gamma_0.
\end{equation}
By expanding $\gamma_{\rm PD}$ in a single-phonon 
limit, i.e.~$e^{\pm\phi(s)}\approx1\pm\phi(s)$ in Eq.~(\ref{gammapd}), the phonon-induced pure-dephasing rate can be approximated by $\gamma_{\rm PD}\approx(\pi/2)J(\Omega_r)\coth{(\Omega_r/2k_BT)}$~\cite{mccutcheon10_2},
which we can further simplify to 
\begin{equation}
\gamma_{\rm PD}(\Omega_r,T) \approx \pi\alpha k_B  T \Omega_r^2,
\label{gammaPD}
\end{equation}
provided that $\Omega_r\ll k_B T,\omega_c$~\cite{ramsay10_2,mccutcheon10_2}. 
Under these constraints, which are satisfied in our experiments, 
the FWHM of the sidebands varies linearly with $\Omega_r^2$, with a gradient
proportional to the exciton-phonon coupling constant $\alpha$. 
Despite the single-phonon expansion, Eq.~(\ref{gammaPD}) still 
captures important non-perturbative phonon effects through the renormalised $\Omega_r$, which is itself 
dependent (non-linearly) upon the coupling strength $\alpha$ and the temperature $T$ [see Eq.~(\ref{Omegar})]. The phonon-induced dephasing rate given by Eq.~(\ref{gammaPD}) is thus distinct from that obtained via a second-order weak-coupling treatment of phonon interactions, in which $\Omega_r$ would be replaced by the bare, unrenormalised $\Omega$~\cite{ramsay10,ramsay10_2}, which is independent of $\alpha$ or $T$. This will be particularly important in describing the temperature dependent behaviour we observe below.

To obtain Fig.~{\ref{driving}} we fit each spectrum to a sum of three Lorenztians to determine  
experimental values for the
positions and widths of the side peaks. Fig.~{\ref{driving}}~(a) shows the extracted renormalised driving strength
as a function of the square root of the laser power, $\sqrt{P}\sim E$,
displaying a linear dependence as expected. In Fig.~{\ref{driving}}~(b) we plot the extracted sideband widths of the blue (centred at $\Omega_r$)
and red (centred at $-\Omega_r$) sidebands 
as a function of $\Omega_r^2$, 
also showing the anticipated 
linear behaviour. Using the FHWM expression given in Eq.~(\ref{fwhm}), we 
obtain values of $\alpha=(2.535\pm0.156)\times10^{-7}~\textup{GHz}^{-2}$ and $\gamma_0=0.135\pm0.062~\textup{GHz}$ from the data in Fig.~{\ref{driving}}~(b). 
The non-zero value of $\gamma_0$ implies that $\Gamma_1/2\Gamma_2 \approx 0.9$ at vanishing driving strength ($\Gamma_1/2\Gamma_2=1$ in the
Fourier transform limit) and may possibly be due to
fluctuating charges trapping in the vicinity of the QD~\cite{robinson00,berthelot06,houel12}.

\begin{figure}[tb]
\centering
\includegraphics[width=0.48\textwidth]{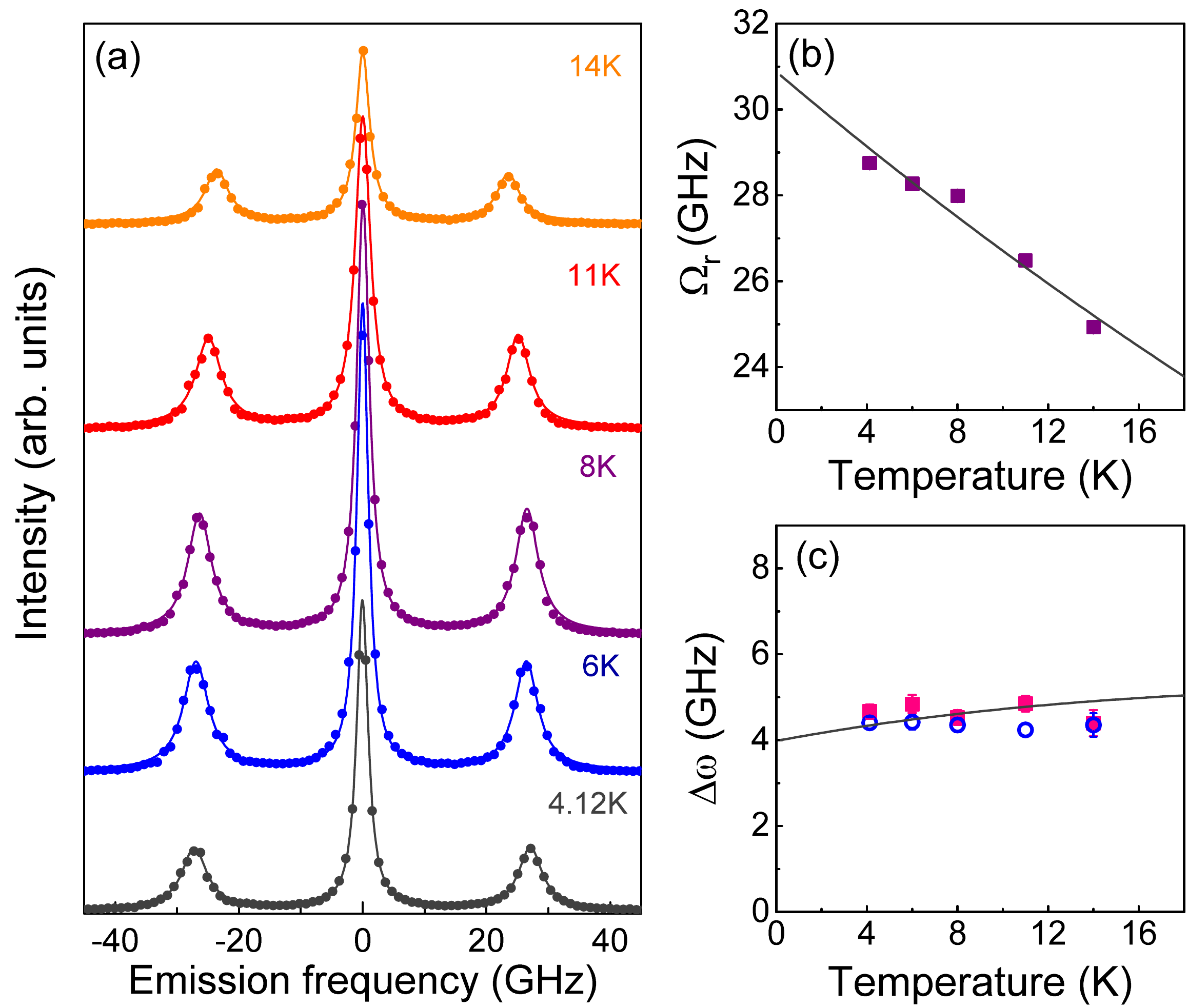}
\caption{(a) High resolution RF spectra of the neutral exciton 
emission at different temperatures. The temperature spans $4-14$~K, and the nominal cw laser power is $9.9$~$\mu$W.
(b) Phonon-renormalised Rabi frequency $\Omega_r$ as a function of temperature.
The data is fitted to Eq.~({\ref{Omegar}}), shown as the solid curve. 
(c) Extracted linewidths of the red (red filled squares)
and blue (blue open circles) sidebands as a function of temperature.
The solid curve shows the expected trend from our non-perturbative 
exciton-phonon coupling theory, using parameters already determined (i.e.~it is not a fit to the data).}\label{temperature}
\end{figure}

We now turn our
attention to the dependence on temperature in the RF spectra.
This was measured 
in the range of $4-14$~K, with the nominal laser power fixed at $9.9$~$\mu$W, as shown in Fig.~\ref{temperature}~(a).
The purple markers in Fig.~\ref{temperature}~(b) show the extracted effective Rabi frequency $\Omega_r$ as a function of temperature. 
The data reveals a clear decrease of
$\Omega_r$ as the temperature is increased. This is 
the most striking evidence of coupling
to phonons in our experiment, since the Rabi frequency 
of a driven atom in the traditional setting of quantum optics 
is determined only 
by the value of the dipole moment and the optical field amplitude, which are both constant in these measurements.

The renormalisation of the Rabi frequency is entirely consistent with our exciton-phonon coupling model. As previously stated, in the regime
in which we are working, the positions of the Mollow sidebands are given approximately by $\pm\Omega_r$, 
as defined in Eq.~({\ref{Omegar}}). From the theory, we expect $\Omega_r$ to decrease with temperature for a fixed driving power, with a slope
determined by the spectral density parameters $\alpha$ and $\omega_c$, which 
together characterise the overall strength of the QD-phonon interaction. Using the extracted value of $\alpha=2.535
\times10^{-7}~\textup{GHz}^{-2}$ from the driving dependent data, we fit
the experimentally measured values of $\Omega_r$ to Eq.~({\ref{Omegar}}) to find a bare Rabi frequency of 
$\Omega=30.88\pm0.52~\textup{GHz}$, 
and a spectral density cut-off frequency of $\omega_c=493.33\pm61.28~\textup{GHz}$. 
We note that this implies $\omega_c\gg\Omega_r$ for
all driving strengths considered previously, thus confirming our assumption that $\exp[-(\Omega_r/\omega_c)^2]\approx 1$ used 
to obtain Eq.~({\ref{gammaPD}}).

To complete the picture we now consider how the sideband widths vary with temperature. The data points plotted in Fig.~{\ref{temperature}}~(c) show
the extracted widths of the blue and red sidebands, while the solid curve shows the expected trend from our theoretical analysis, for which the
FWHM is given by Eq.~(\ref{fwhm}) using $\gamma_{\rm PD}$ as defined in Eq.~(\ref{gammaPD}).  
We note that having fully determined all parameters from other measurements, there is no fitting procedure used to obtain this curve. Intriguingly, we
see that theoretically we expect a sub-linear increase in the sideband widths across this temperature range, 
which also appears to be supported by the data. From
Eq.~({\ref{gammaPD}}), the phonon-induced dephasing rate depends on the product $\alpha\,T\,\Omega_r^2$. 
The explicit dependence on $T$ is a first order process, reflecting the increase in the occupation
number of phonons with a frequency corresponding to the Rabi energy. The implicit dependence on $T$ through $\Omega_r$, 
however, is a non-perturbative effect describing the renormalisation of the dressed state energy splitting which serves to reduce the 
dephasing rate. A competition between these two processes gives rise to
an overall weak dependence on temperature for the parameters we have determined here. 

We stress again that this effect cannot be captured
by a perturbative weak-coupling treatment, in which $\Omega_r\to\Omega$ independent of temperature. 
In such a case the dephasing rate, and hence
the sideband widths, would increase linearly with 
temperature~\cite{ramsay10,ramsay10_2,monniello13,mccutcheon10_2}. 
However, it is important to note that the non-linear temperature dependence we 
observe is not at odds with the linear dependence seen in Refs.~\cite{ramsay10,ramsay10_2,monniello13}. 
The difference in behaviour arises as a result of the non-perturbative regime in which our experimentally 
determined parameters place us, and thus depends sensitively on the phonon coupling characteristics of the individual QD sample studied. 
Specifically, we extract a larger value of the coupling strength $\alpha$ than in Refs.~\cite{ramsay10,ramsay10_2,monniello13} and also a smaller 
cut-off frequency $\omega_c$, the latter of which suggests a larger QD size due to the inverse relationship between the two~\cite{ramsay10,ramsay10_2}. The combination of a larger $\alpha$ and smaller $\omega_c$ ensures that the renormalisation of the Rabi frequency, which ultimately gives rise to the non-linear temperature dependence  
[see Eq.~(\ref{Omegar})], is enhanced in our experiments.

In summary, we have presented the first systematic 
exploration of temperature dependence in the RF emission spectra of a solid-state QD emitter.
Our experiments and theoretical description were first calibrated through an analysis of the dependence on driving strength,
in which we confirmed the expected increase in sideband splitting and linewidth with increasing driving strength. 
We then showed a systematic {\it decrease} of sideband splitting with increasing temperature at constant driving, an effect we attribute to a phonon-induced
temperature-dependent suppression of the effective Rabi frequency $\Omega_r$~\cite{mccutcheon10_2}. 
Resulting from this renormalisation, we observed evidence of non-perturbative phonon coupling effects in the insensitivity of sideband 
linewidths to an increase in temperature. This suggests that changes in temperature may not be as detrimental to 
QD coherence as a weak-coupling analysis would otherwise indicate, which could have important repercussions 
in the application of QDs to future quantum technologies---such as high quality single photon generation in strong cavity-QD 
coupling systems~\cite{strongcoupling} and QD-based thermometry~\cite{thermometry}---as well as for other solid-state emitters~\cite{molecularRF,superconductingRF,diamondg2}.

\textit{Acknowledgements}: We thank C. M. Chen and Z. G. Lu for helpful discussions. 
This work was supported by the National Natural Science Foundation of China, the Chinese Academy of Sciences, 
the National Fundamental Research Program (under Grant No: 2011CB921300, 2013CB933300), the State of Bavaria, 
by the BMBF and EPSRC within the CHIST-ERA project SSQN, and 
project SIQUTE (contract EXL02) of the 
European Metrology Research Programme (EMRP). 
The EMRP is jointly funded by the EMRP participating countries within EURAMET and the European Union. 
S. H. acknowledges the CAS visiting professorship. D. P. S. M. thanks 
CONICET and A. N. thanks Imperial College London and the University of Manchester for financial support.


\begin{thebibliography}{10}
\bibitem{note}
To ensure that this corresponds to the temperature of the QD itself, we have performed COMSOL simulations  
from which we estimate that the excitation laser heats the sample by no more than $10^{-6}~\mathrm{K}$.
\end{thebibliography}
\end{document}